\documentclass[prd,twocolumn,nofootinbib]{revtex4}
\usepackage{graphicx} 
\usepackage{lipsum} 
\usepackage{amsmath}
\usepackage{subcaption}
\usepackage{siunitx}
\usepackage{svg}
\usepackage{url}

\begin{document}

\title{All-polarisation beamsplitters for interferometer applications}
\author{S.L. Kranzhoff$^{1,2}$, Z. Van Ranst$^{1,2}$, J. De Bolle$^{3}$, S. Coessens$^{3,5}$, S.L. Danilishin$^{1,2}$, C. Detavernier$^{3}$, P.F. Smet$^{4}$, A.P. Spencer$^{6}$, J. Steinlechner$^{1,2}$, S. Steinlechner$^{1,2}$, M. Vardaro$^{1,2}$, S. Hild$^{1,2}$}
\affiliation{$^{1}$Maastricht University, Department of Gravitational Waves and Fundamental Physics, 6200MD Maastricht, the Netherlands}
\affiliation{$^{2}$Nikhef, Science Park 105, 1098 XG Amsterdam, the Netherlands}
\affiliation{$^{3}$Department of Solid State Sciences,
CoCooN, Ghent University, 9000 Gent, Belgium}
\affiliation{$^{4}$Department of Solid State Sciences,
LumiLab, Ghent University, 9000 Gent, Belgium}
\affiliation{$^{5}$Department of Materials, Textiles and Chemical Engineering, SMS, Ghent University, 9052 Zwijnaarde, Belgium}
\affiliation{$^{6}$SUPA, Institute for Gravitational Research, University of Glasgow, Glasgow G12 8QQ, United Kingdom}

\begin{abstract}
Optical beamsplitters with similar properties for orthogonal, linear polarisation modes are required for realising polarisation-based speedmeter schemes to reduce back-action noise in gravitational-wave interferometers. In this paper, we investigate two beamsplitter coatings obtained from Laseroptik GmbH and Optoman on a best-effort basis that aim for a 50/50 power splitting ratio and equal overall phase shift for two orthogonal, linear polarisation modes interacting with the optic. We show that while Laseroptik GmbH opted for coating stack with 22 alternating layers of Ta$_{2}$O$_{5}$ and SiO$_{2}$, Optoman produced a much thinner coating made of 5 SiO$_{2}$ and SiO$_{x}$ ($0 < x < 2$) layers. With these strategies, the Laseroptik coating achieves an equal power reflectivity of $51\,\%$ at $46^{\circ}$ angle of incidence, and zero phase shift between both polarisations at $44.25^{\circ}$ angle of incidence. The Optoman coating achieves power reflectivities of $49\,\%$ for s-polarisation and $51\,\%$ for p-polarisation with a differential phase shift around $5^{\circ}$ largely independent of the angle of incidence.
\end{abstract}
\maketitle

\section{Introduction}
Dual-recycled Michelson interferometer topologies, like Advanced LIGO \cite{aLIGO2015} and Advanced Virgo \cite{aVirgo2015}, have been very successful in detecting gravitational waves from black hole and neutron star mergers \cite{GWTC2019,GWTC2024,GWTC2023}. These instruments rely on a central optical beamsplitter component that splits the laser beam into two orthogonal arms and recombines the back-travelling beams from both arms to measure arm length changes.

Over a wide frequency range, the design sensitivity of current and future gravitational-wave detectors is fundamentally limited by quantum noise \cite{Danilishin2019}, which originates from the particle-like nature of light and its interaction with suspended optics and photodetectors in the interferometer. One way to overcome this semi-classical sensitivity limit are speedmeters \cite{Braginsky1990}, which measure a quantum non-demolition observable and thereby allow significant reduction of radiation pressure noise, the manifestation of quantum noise limiting the interferometer sensitivity at low frequencies. Several incarnations of speedmeters have been proposed \cite{Braginsky2000, Purdue2002, Chen2003, Wade2012, Wang2013, Knyazev2018, Danilishin2019}, of which polarisation-based speedmeters are favourable because they require the least infrastructure changes to operating interferometers.

Polarisation-based speedmeters rely on two orthogonal polarisation modes that coherently probe the mirror position independently at different times to achieve a mirror speed measurement. This requires the interferometer optics to have the same optical properties for both modes. According to Fresnel equations, reflectance, transmittance and phase shift acquired by light impinging under non-normal incidence at an interface are generally different for p- and s-polarisation.  Symmetric beamsplitters are commonly designed for $45^{\circ}$ angle of incidence and feature two coated surfaces, one for achieving the 50:50 power splitting ratio between reflected and transmitted beams, and one anti-reflective coating to reduce reflection from the back surface of the optic to a few hundreds of ppm (parts per million). 

Coatings are typically composed of multiple interfaces of alternating high- and low-refractive index dielectric materials. Careful choice of materials, number of layers and individual layer thickness have been shown previously to lead to coating designs with similar properties for p- and s-polarised light, for instance, by using a graded index film~\cite{Zukic1988} or more than two materials~\cite{Mordechai1992}. Refractive gratings and metasurfaces have also been investigated as alternative design approaches \cite{Feng2009, Shen2022}.

Coatings with similar properties for all polarisations are not only interesting for the feasibility of polarisation-based speedmeters, but have also been investigated among others in the context of photopolarimetry~\cite{Azzam2003}, optical communications~\cite{Zhang2006} and space applications~\cite{Rothhardt2023}. 
In this paper, we present the characterisation of two different all-polarisation coatings provided by Laseroptik GmbH \footnote{Laseroptik GmbH, Johannes-Ebert-Str. 1, 30826 Garbsen, Germany - \url{https://www.laseroptik.com/}} and Optoman \footnote{Optoman, Ukmerges str. 427, LT-14185 Vilnius, Lithuania - \url{https://www.optoman.com/}} on a best-effort basis.  A total of 21 beamsplitter samples were characterised: 8 Laseroptik samples (four plane and four wedged) and 13 Optoman samples (6 plane and 7 wedged). We will discuss the coating design strategies in section~\ref{subsub:decomp}, analyse the achievable power-splitting ratio symmetry in section~\ref{subsec:power-split} and phase shifts introduced between both polarisations in section~\ref{subsec:phase-shift}.

\section{Measurements}
\subsection{Coating decomposition} 
\label{subsub:decomp}
Before assessing the functional characterisation of the two coatings, their structure and composition are studied to gain a better understanding of the design strategies chosen by the two manufacturers. To this end, a combination of scanning electron microscopy (SEM) and energy-dispersive X-ray spectroscopy (EDX) was employed to investigate the cross-sectional morphology and elemental distribution of the coatings.

A sample of each coating, deposited on a fused silica substrate, was cut with a water-cooled diamond saw, after which a cross-sectional surface, perpendicular to the coating surface, was prepared for SEM analysis. This was done by first mechanically grinding the surface up to 4000 grit using SiC foil (Struers), and subsequently polishing the samples using OP-U 0.04 µm standard colloidal silica suspension in the final polishing step.

After polishing, the samples were mounted in the SEM. Due to the insulating fused silica substrates, charging effects complicate proper visualisation of the coatings in an SEM operating under ultra-high vacuum conditions. Furthermore, drifting of the electron beam makes EDX line scans unreliable. Therefore, a variable pressure SEM (VPSEM)~\cite{goldstein2017}, which can operate at an elevated sample chamber pressure, was employed, allowing for better control over the sample's electrostatic charging. For this study, a Hitachi S-3400N VPSEM was used, operating at a sample chamber pressure of 20 Pa. The surface charge on the samples is removed by ionization of the residual gas in the sample chamber. Because a VPSEM is not compatible with secondary electron (SE) detectors, imaging in this setup is done with backscattered electrons (BSE). Although using BSE instead of SE leads to a lower resolution, BSE images feature a large atomic, or Z-, contrast. For EDX analysis, the setup is equipped with a ThermoScientific UltraDry detector and NORAN System 7. For imaging with secondary electrons, a JEOL JSM-IT800 SEM operating in ultra-high vacuum was used and, despite charging effects, some images of the coatings could be made. All accelerating voltages used were chosen to optimise the sensitivity to the relevant elements and image quality, while minimising background.

\begin{figure}[htbp]
\centering
\begin{subfigure}[b]{\linewidth} 
    \includegraphics[width=0.625\linewidth]{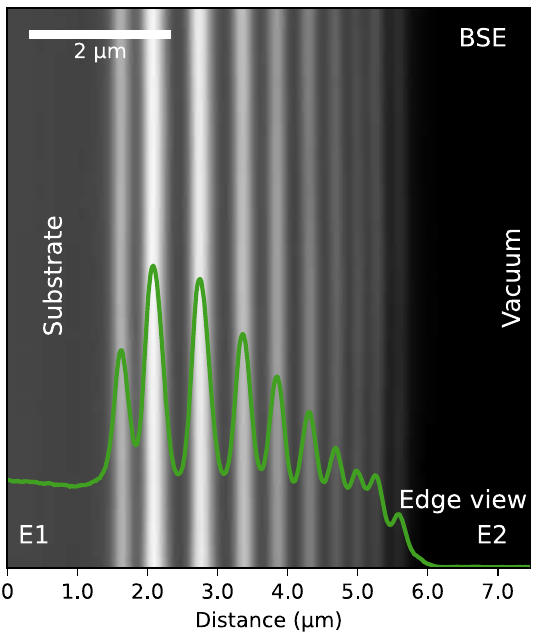}
    \caption{}
    \label{LOP_images:a}
\end{subfigure}\\
\begin{subfigure}[b]{\linewidth} 
    \includegraphics[width=0.625\linewidth]{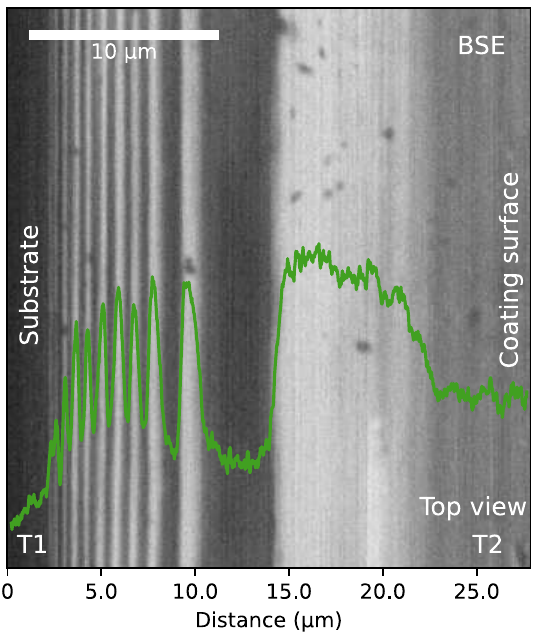}
    \caption{}
    \label{LOP_images:b}
\end{subfigure}\\
\begin{subfigure}[b]{\linewidth} 
    \includegraphics[width=0.625\linewidth]{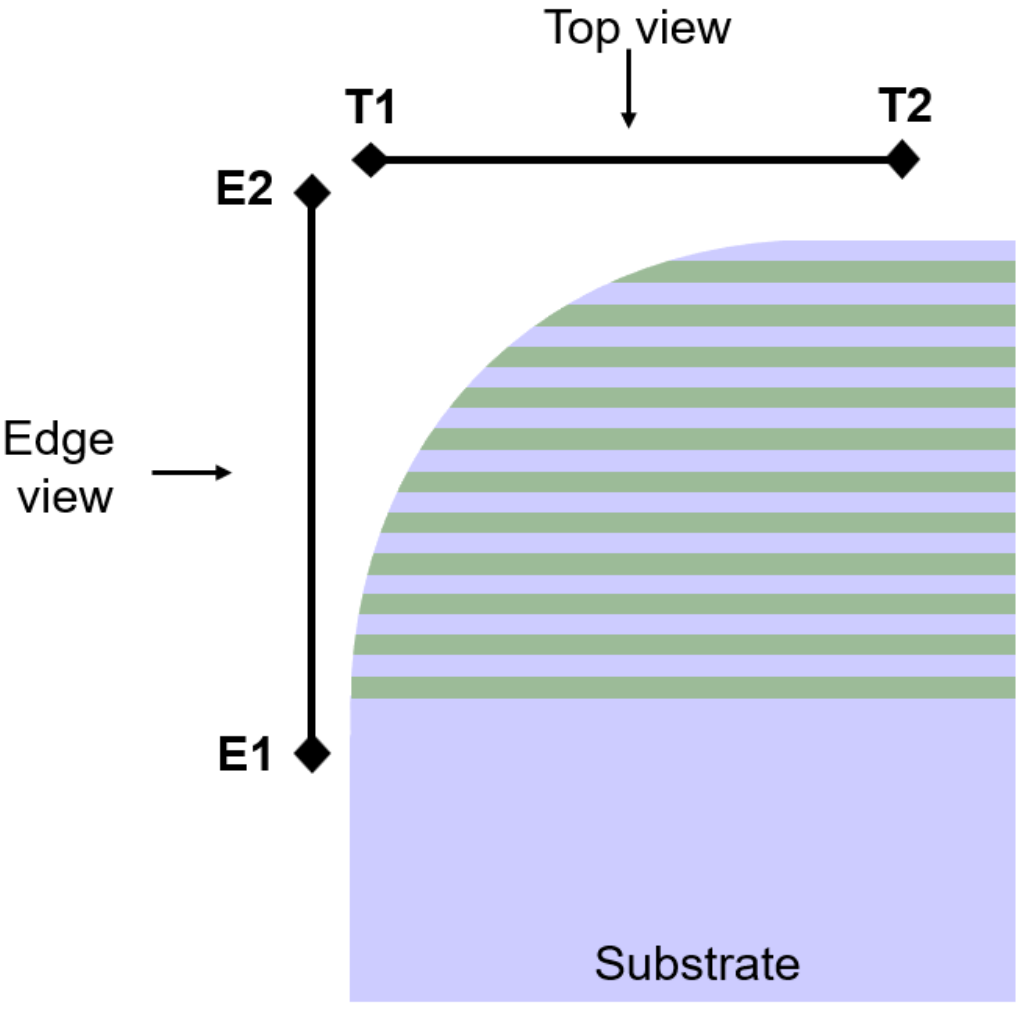}
    \caption{}
    \label{LOP_images:c}
\end{subfigure}
\caption{(a) BSE image of the Laseroptik coating acquired in the VPSEM, looking at the polished cross-section (edge view). (b) BSE image of the Laseroptik coating acquired in the VPSEM, looking at the coating surface near the polished edge of the sample (top view).
The green lines represent the vertically integrated image intensity. (c) A schematic illustration of how the measurements were performed to help with image interpretation.}
\label{LOP_SEM_images}
\end{figure}

\begin{figure}[htbp]
    \includegraphics[width=0.625\linewidth]{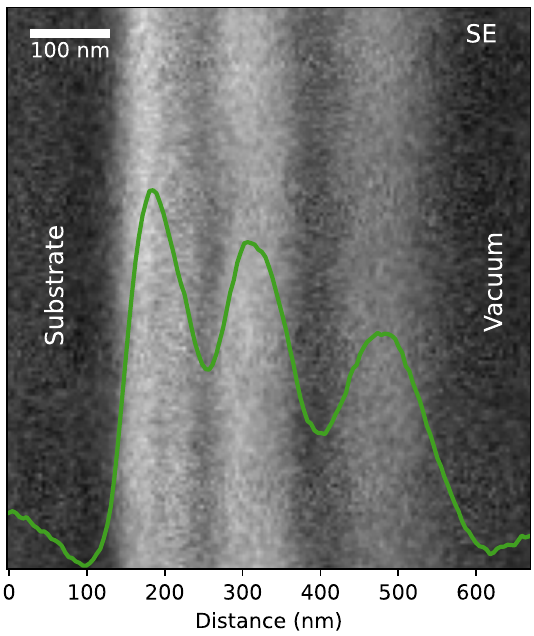}
    \vspace{-0.2cm}
    \label{OMP_image}
\caption{SE image of the Optoman coating, acquired in the JEOL SEM. The green line represent the vertically integrated image intensity.}
\label{OMP_SEM_images}
\end{figure}
Figs.~\ref{LOP_images:a} and~\ref{LOP_images:b} show BSE images of the Laseroptik beamsplitter coating from different perspectives. To help with image interpretation, fig.~\ref{LOP_images:c} provides a schematic illustrating how the measurements were performed. Fig.~\ref{LOP_images:a} presents a cross-sectional (edge view) BSE image of the coating, acquired in the VPSEM with an accelerating voltage of 15 kV. The green line shows the pixel intensity integrated along the direction parallel to the coating layers (i.e. vertically in the image). A strong contrast between bright and dark layers can be observed, indicating that the coating is made of alternating layers with high chemical contrast, where the bright layers have a higher average atomic number than the dark layers. A total of 10 bright and 9 dark layers can be counted. The layers increase in thickness from the surface towards the substrate, except for the bottom layer, which is again thinner. Fig.~\ref{LOP_images:b} shows a top-view BSE image of the coating, acquired near the cut and polished edge of the sample in the VPSEM with an accelerating voltage of 6 kV. The green curve again represents the vertically integrated pixel intensity. The visibility of the individual coating layers in this top view indicates that the polishing was not perfect: the polished surface is slightly sloped, as schematically represented in fig.~\ref{LOP_images:c}. This explains the decrease in image intensity towards the coating surface in fig.~\ref{LOP_images:a}. The surface layer appears with lower intensity, indicating it is composed of a material with a lower average atomic number. From this image, 11 bright layers and 11 dark layers can be counted. Consequently, the Laseroptik beamsplitter coating consists of 22 layers in total. Although the top three layers of the coating cannot be distinguished from fig.~\ref{LOP_images:a} due to loss of image intensity and contrast, this image is more representative for estimating the coating thickness and the individual layer thicknesses.

A cross-sectional BSE image of the Optoman beamsplitter coating, acquired in the VPSEM, did not reveal any structural information due to a lack of chemical contrast. This suggests that the coating layers are made of similar materials with comparable average atomic numbers. Despite charging effects, an SE image of the coating could be obtained under ultra-high vacuum conditions in the JEOL SEM, using an accelerating voltage of 5 kV (fig.~\ref{OMP_SEM_images}). The green line again shows the vertically integrated pixel intensity. A total of 5 layers can be identified.

\begin{figure}[htbp]
\centering
\begin{subfigure}[b]{\linewidth} 
    \includegraphics[width=\linewidth]{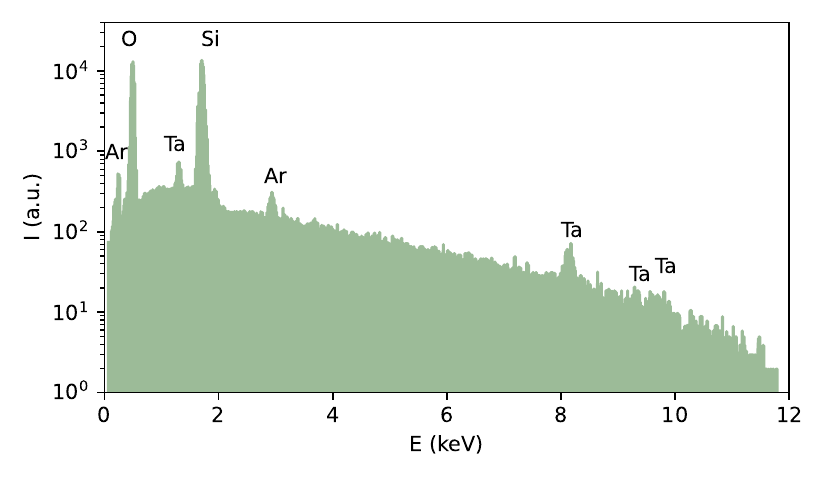}
    \caption{}
    \label{EDX_spectra:a}
\end{subfigure}\\
\begin{subfigure}[b]{\linewidth}
    \includegraphics[width=\linewidth]{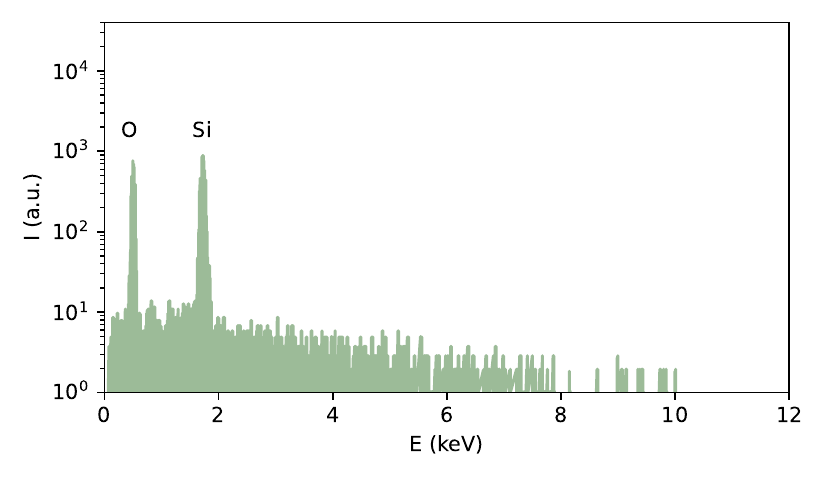}
    \caption{}
    \label{EDX_spectra:b}
\end{subfigure}
\caption{EDX spectra measured on cross-sections of the beamsplitter coatings, acquired with the ThermoScientific UltraDry detector and
NORAN System 7 on the VPSEM. (a) The coating from Laseroptik only shows peaks from oxygen (O), silicon (Si), tantalum (Ta) and argon (Ar). The coating is therefore made of a stack of alternating layers of Ta$_2$O$_5$ and SiO$_2$, with the argon likely being contamination from the deposition process. (b) The coating from Optoman only shows peaks of oxygen (O) and silicon (Si). It is likely that the coating is a stack of SiO$_2$ layers and SiO$_x$ layers, where the SiO$_x$ is either amorphous silicon ($x=0$) or substoichiometric silicon oxide ($0<x<2$).}
\label{EDX_spectra}
\end{figure}

\begin{figure*}[t]
\centering
\begin{subfigure}[b]{0.67\textwidth}
    \centering
    \includegraphics[height=4.5cm]{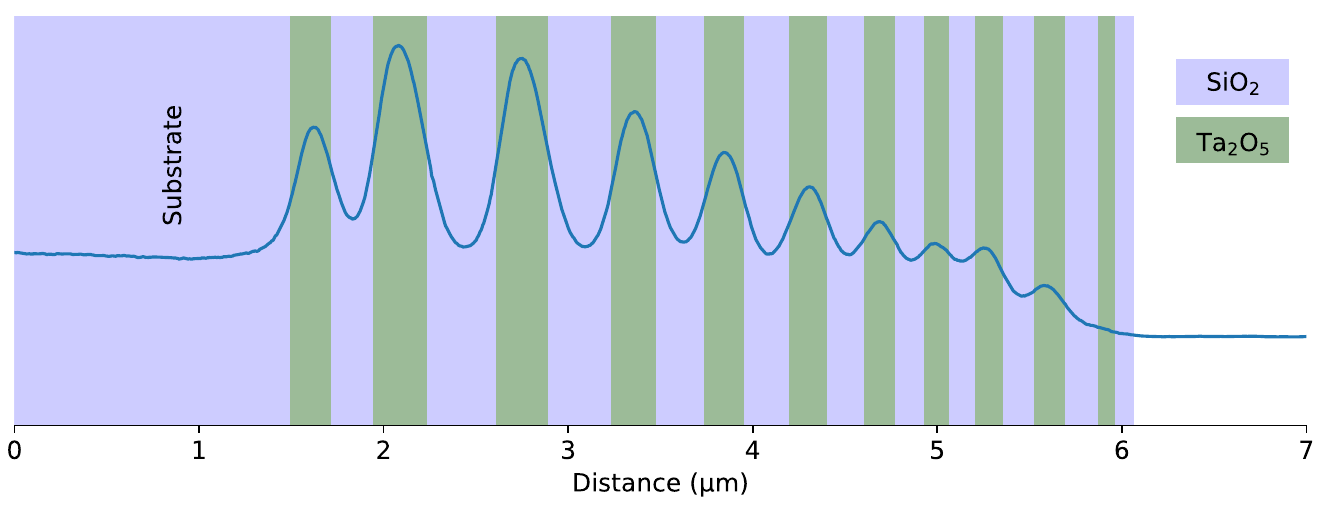}
    \caption{}
    \label{coating_result:a}
\end{subfigure}
\begin{subfigure}[b]{0.32\textwidth}
    \centering
    \includegraphics[height=4.5cm]{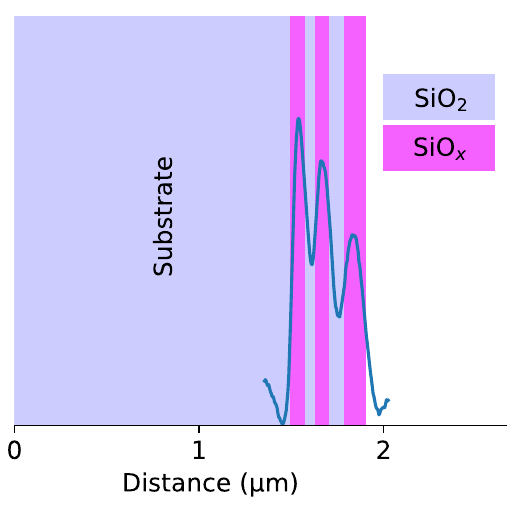}
    \caption{}
    \label{coating_result:b}
\end{subfigure}
\caption{Visual representation of the beamsplitter coating composition and structure (with estimated layer thicknesses) as obtained from the analysis of the SEM images and EDX spectra, for (a) the Laseroptik coating and (b) the Optoman coating. The horizontal scale of both figures is the same to exemplify the strong difference in coating design between both manufacturers.}
\label{coating_result}
\end{figure*}
Fig.~\ref{EDX_spectra} shows the EDX spectra of the two investigated beamsplitter coatings, measured in cross-section in the VPSEM. For the Laseroptik coating, an accelerating voltage of $12$ kV was used. From the measured spectrum (fig.~\ref{EDX_spectra:a}), it can be observed that only oxygen, silicon, tantalum and argon were detected. Together with the BSE images (figs.~\ref{LOP_images:a} and~\ref{LOP_images:b}), this indicates that the coating is made of a stack of alternating layers of Ta$_2$O$_5$ and SiO$_2$. An EDX linescan confirms that the bright layers have a higher Ta-content, while the dark layers have a higher Si-content. The argon is most likely contamination from the coating deposition. For the Optoman coating, an accelerating voltage of 10 kV was used. From the measured spectrum (fig.~\ref{EDX_spectra:b}), only oxygen and silicon can be observed. Since the attempt to obtain a BSE image of this coating was unsuccessful due to insufficient chemical contrast, it is likely that the coating is made of a stack of SiO$_2$ layers and SiO$_x$ layers, where SiO$_x$ is either amorphous silicon ($x=0$) or substoichiometric silicon oxide ($0<x<2$)~\cite{song2008}.

\indent Figs.~\ref{coating_result:a} and ~\ref{coating_result:b} represent the estimated structure and composition of the coatings as obtained from the SEM and EDX analysis. The blue curves are the vertically integrated image intensities shown in figs.~\ref{LOP_images:a} and~\ref{OMP_SEM_images}. Local thresholding was employed to estimate the position of layer interfaces. The horizontal scale of figs.~\ref{coating_result:a} and~\ref{coating_result:b} is the same to highlight the pronounced difference in coating design between the two manufacturers. Based on these results, the total coating thickness is around $5$ \si{\micro\meter} for the Laseroptik coating and $400$ nm for the Optoman coating. A measurement of the total coating thickness was also performed with a Taylor Hobson TalyStep profilometer. A differential measurement between the middle of the sample surface and the uncoated edge resulted in $5\pm1$ \si{\micro\meter} for the Laseroptik coating and $0.5\pm0.1$ \si{\micro\meter} for the Optoman coating, confirming that the SEM analysis produced values of the correct order of magnitude.

\indent The analysis presented here reveals notable differences between the two coatings, in both their layer structure and material composition. The results indicate that the two manufacturers employ markedly different coating designs and manufacturing approaches.

\subsection{Power-splitting ratio} 
\label{subsec:power-split}
\begin{figure}[b]
    \centering
    \includegraphics[width=0.85\linewidth]{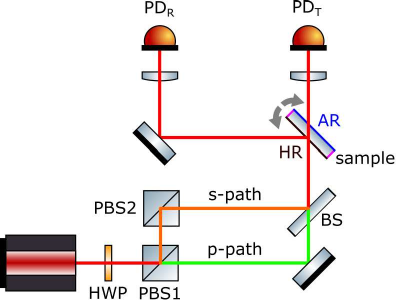}
    \caption{Schematic of the setup used to measure the power splitting ratio for the all-polarisation beam splitter under test (labeled sample).}
    \label{BS_char_paper}
\end{figure}

The power splitting ratio of the beamsplitter is measured with the setup shown in fig.~\ref{BS_char_paper}. The laser beam, with a wavelength $\lambda = 1550$nm, is initially linearly polarised and incident on a half-wave plate (HWP). The HWP is used to tune the polarisation direction of the laser beam that enters a first polarising beam splitter (PBS). This beam splitter, labeled as PBS1, separates the laser beam into the two orthogonal s- and p-polarisation modes. This split-off is necessary to probe the beamsplitter sample with both polarisations independently. Since the polarisation purity of the transmitted beam through the PBS is generally better than that of the reflected beam (transmission p-pol $> 97\%$ vs reflection s-pol $> 99.5\%$), a second PBS (PBS2) is used to filter out any remaining p-polarised light from the mainly s-polarised beam in reflection. Assuming that the incoming beam is $45^{\circ}$ linearly polarized, the reflected beam will contain a maximum of 3\% p-polarised light. This is reduced to $0.09\%$ after passing PBS2. Both of the p- and s-polarised beams are recombined at a 50/50 beam splitter (BS), where one of the output ports sends the combined beam to the all-polarisation beamsplitter that is to be characterised. The other output is dumped. The all-polarisation beamsplitter sample is placed on a precision rotation mount to allow for different angles of incidence. The outgoing beams are each focused on a photodiode (PD) by a lens. The only difference is that the transmitted beam is directly captured by PD\textsubscript{T}, while the reflected beam first passes a steering mirror before being captured by PD\textsubscript{R}. 

The probed beamsplitter is initially put under a $45^{\circ}$ angle of incidence with respect to the p-polarised beam, after which BS and PBS2 are adjusted to co-align the s-polarised beam onto the first one. This ensures that both beams hit the same spot on the all-polarisation beamsplitter coating and the photodiodes. 

Each measurement starts with the sample at an angle of $2^{\circ}$ away from its initial position. This is the largest possible angle for the beamsplitter such that the beam does not clip on the steering mirror in reflection. First, one of the polarised beam paths is blocked, so that the reflectance and transmittance for the orthogonal polarisation can be measured by PD\textsubscript{R} and PD\textsubscript{T} respectively. After that, the other path is blocked so that the same measurement can be realised for the other polarisation. 

\begin{figure}[htbp]
\centering
\hspace*{-10mm}  
\begin{subfigure}[b]{\linewidth} 
    \includegraphics[width=1.1\linewidth]{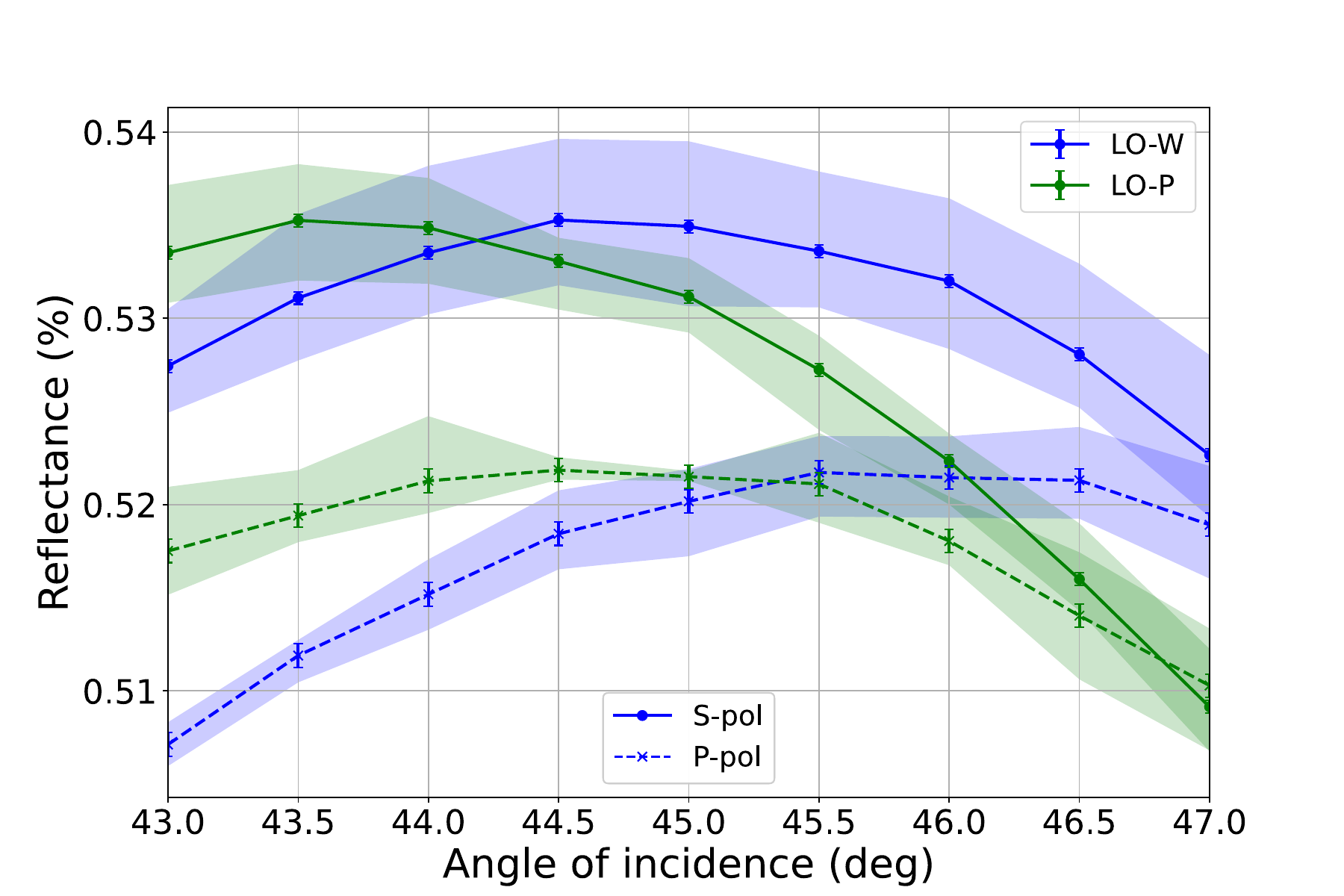}
    \caption{}
    \label{reflectance_avg_LO}
\end{subfigure}\\
\hspace*{-10mm} 
\begin{subfigure}[b]{\linewidth}
    \includegraphics[width=1.1\linewidth]{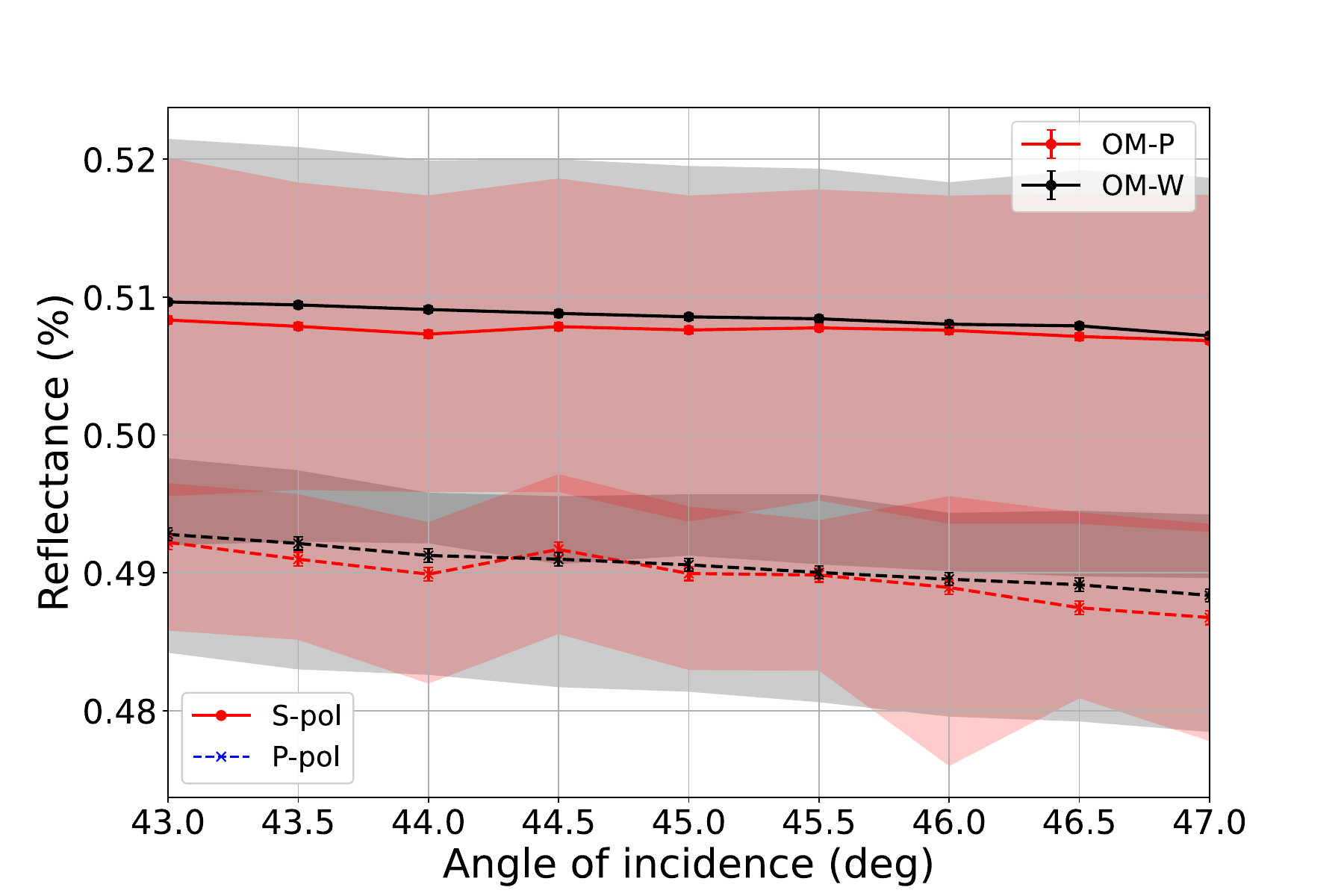}
    \caption{}
    \label{reflectance_avg_OM}
\end{subfigure}
\caption{Average reflectance of the all-polarisation beamsplitter coatings produced by Laseroptik (a) and Optoman (b). The additions P and W indicate whether the substrate of the beamsplitter is plane or wedged, respectively. The shaded areas are created by the highest and lowest reflectance that was measured for each category of samples, to indicate the variations for each coating.}
\label{reflectance_avg}
\end{figure}

Next, the beamsplitter is rotated by $0.5^{\circ}$ towards the central $45^{\circ}$ position and the reflectance and transmittance are measured again. This changes the beam path of the reflected beam, so the steering mirror is used to align the beam back onto PD\textsubscript{R}. This procedure is repeated until the entire $\pm 2^{\circ}$ range around the nominal $45^{\circ}$ angle of incidence is covered. The seventh plane Optoman beamsplitter was used as BS in fig.~\ref{BS_char_paper}. The wedged samples all have a wedge of $2^{\circ}$ on the side with the antireflective coating. This sample is orientated in such a way that the width of the wedge increases horizontally, away from the direction of the transmitted beam. So also the refraction of the beam is only affected in the horizontal direction. 

The reflectance is calculated by taking the reflected power $P_R$ and dividing it by the total power:

\begin{equation}
    R = \frac{P_R}{P_R + P_T}\,,
\end{equation}

where $P_T$ is the transmitted power. This ensures that the measurement is insensitive to general power fluctuations. The optical power is inferred from the calibrated voltage output given by each photodiode. The calibration factors (in terms of output voltage per unit input power) are determined by varying the power input for a photodiode and making a linear fit of the resulting voltage output, using a least squares fitting algorithm.

The statistical uncertainties on the reflectance are obtained by repeating a measurement of one of the samples (a plane Optoman sample) and calculating the sample standard variation for each angle and polarisation. Then, the standard deviation is averaged across all angles and assigned equally to each data point. This measure of uncertainty is displayed by the error bars of each measurement in fig.~\ref{reflectance_avg} and indicates statistical errors within that measurement (like the beam hitting slightly different spots on the photodiode or the finite precision of the rotation mount). The maximal variations \textit{across different samples from the same type} are represented by the shaded areas in fig.~\ref{reflectance_avg}. This is a measure of the physical differences between individual coating samples, but also includes statistical uncertainties caused by different positioning in the mount, for example.

The averaged values of the measured reflectances are shown in fig.~\ref{reflectance_avg}. It follows from these measurements that the Laseroptik samples show more angular dependence than the ones from Optoman, likely because they use a more complex coating structure with more layers. The less-layered Optoman coating shows a very consistent performance in terms of reflectance as a function of incident angle, with an averaged reflectance of ${\sim}51\%$ for s-polarised light and ${\sim}49\%$ for p-polarised light. Reflectances between individual coatings can vary up to 3\% for s-polarised light and up to 2\% for p-polarised light. Interestingly, the coatings average ${\sim} 2\%$ differences in reflectance between s- and p-polarisations across the entire $5^{\circ}$ angular sweep.

The Laseroptik samples show a point where the reflectance of s- and p-polarisations are equal. However, this condition is met close to $47^{\circ}$ at 51\% reflectance for the plane beamsplitter samples (LO-P). Variations in reflectance are limited for all Laseroptik samples, with $<0.9\%$ difference between individual samples at any given angle of incidence. 

There exist, sometimes significant, deviations between the reflectance curves of different samples. This could be because of inherent differences between the coatings from the same run, or other systematic uncertainties that come from exchanging the sample in the experiment.

\subsection{Dark fringe offset} 
\label{subsec:phase-shift}
\begin{figure}
    \centering
    \includegraphics[width=\linewidth]{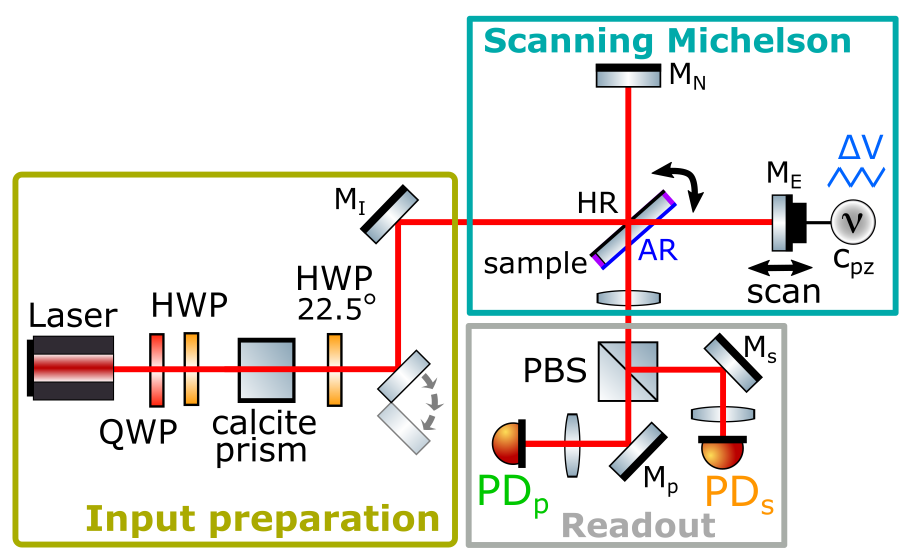}
    \caption{Schematic of the scanning Michelson interferometer setup used to measure the phase shift introduced between p- and s-polarisation modes introcuded by the beamsplitter sample.}
    \label{fig:smi-setup}
\end{figure}

The overall phase shift introduced by the beamsplitter was measured in a scanning Michelson interferometer setup, see fig.~\ref{fig:smi-setup}. The polarisation of the input light (from the left) was prepared using a half-wave plate, a quarter-wave plate and a high-extinction coefficient ($\epsilon\sim 10^{-6}$) calcite prism. The input polarisation impinging on the beamsplitter sample was chosen to be $45^{\circ}$-linear, i.e. with equal power in the p- and s-polarisation modes. The beamsplitter to be characterised was placed in the center of the scanning Michelson interferometer on a rotation mount to allow for different angles of incidence. The rotation stage was calibrated to $45.00^{\circ}\pm 0.25^{\circ}$ angle of incidence by observing the reflection of the beamsplitter in the far field. The end mirror in transmission (M$_{\text{E}}$) was aligned to overlap incoming and reflected light. This mirror was piezo-driven and remained unchanged after initial alignment. The end mirror in reflection of the beamsplitter (M$_{\text{N}}$) needed to be realigned for each angle of incidence to overlap the beams from both arms on the beamsplitter. Due to the geometry of the optics (one-inch diameter) and the arm length of the Michelson (15\,cm), the measurable range was limited to $\pm 2^{\circ}$ around the nominal $45^{\circ}$ angle of incidence. At the dark port of the Michelson interferometer, the laser was split into its p-polarisation (transmission) and s-polarisation (reflection) components using a PBS and detected by individual PDs. 

At the start of each measurement, the end mirror in reflection of the main beamsplitter (M$_{\text{N}}$) was aligned to overlap the beams from both arms of the Michelson interferometer, and the steering mirrors downstream of the PBS (M$_{\text{p,s}}$) were used to maximise the signals on both PDs. A triangular ramp of 234\,Hz was applied to scan the relative arm length of the interferometer over multiple fringes per upward and downward ramp. Fine adjustments of the end mirror (M$_{\text{N}}$) were used to maximise the fringe contrast, before the data was saved. 

\begin{figure}
    \centering
    \includegraphics[width=\linewidth]{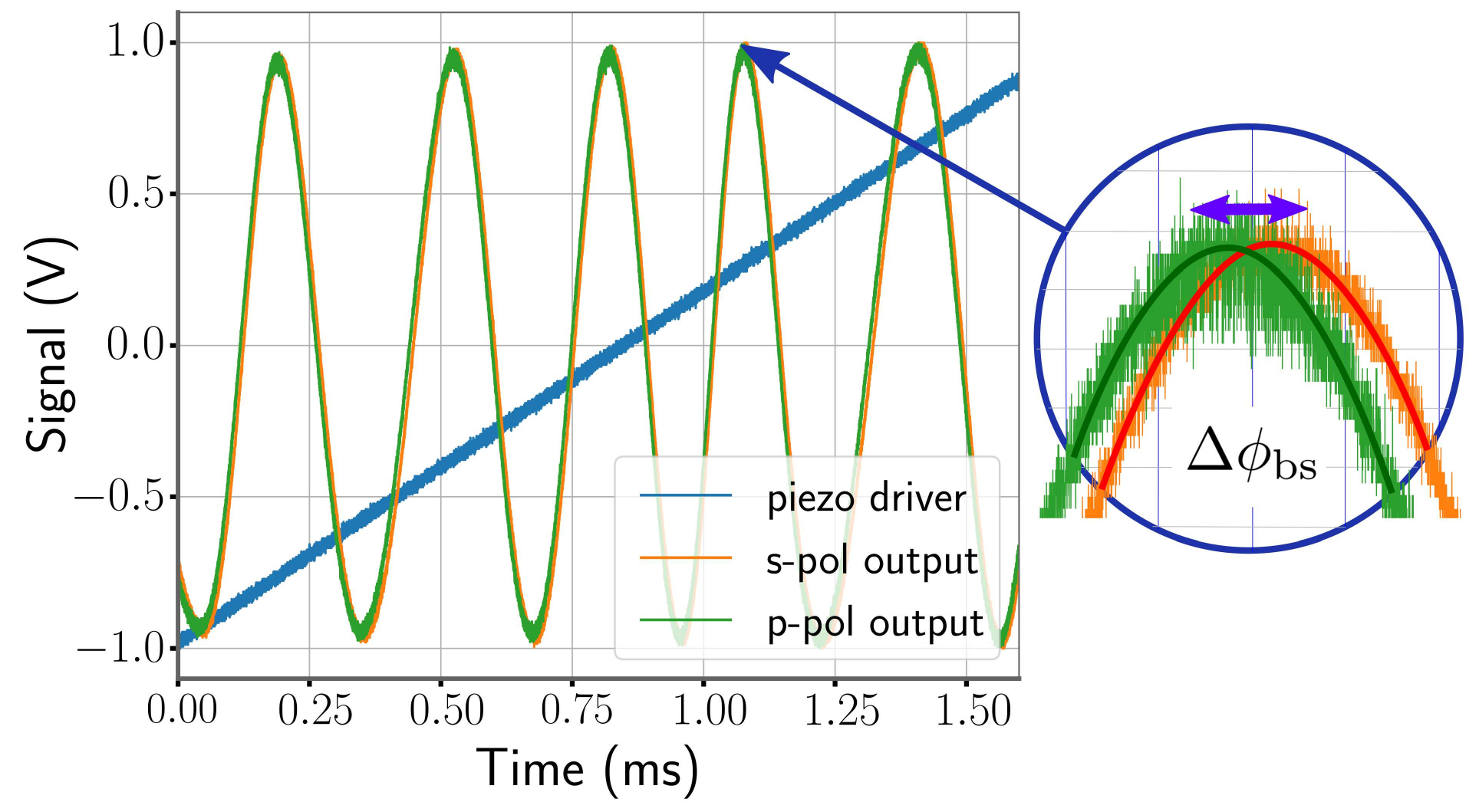}
    \caption{Signals from the scanning Michelson experiment averaged over all upward ramps.}
    \label{fig:osci-sig}
\end{figure}

In a first post-processing step, fringe signals for both polarisations were cut at the turning points of the triangular piezo ramp and averaged separately between data streams corresponding to upward and downward ramps. Fig.~\ref{fig:osci-sig} shows an example of an averaged signal for upward ramps. The PD signals measured for both polarisations $i=$p,s are generally of the form 
\begin{align}
    V_{i}=A_{i}\cdot\cos^2\left[\frac{\pi c_{\text{pz}}\Delta V}{\lambda}+\frac{\Delta\phi_{\text{bs},i}}{2}\right]
    \,,
\end{align}
where the fringe amplitude $A_{i}$ is dependent on the laser input power, the beamsplitter power reflectivity and the responsivity of the individual photodetectors. The Michelson arm length change depends on the length coupling $c_{\text{pz}}$(m/V) per voltage change $\Delta V$, which is potentially nonlinear due to piezo hysteresis and calibrated to the laser wavelength $\lambda=1550\,$nm. The phase shift $\Delta\phi_{\text{bs}}$ between the fringe signals was the quantity to be measured. The phase of the p-polarisation fringe signal was taken as reference ($\Delta\phi_{\text{bs,p}}=0$) and the sign of $\Delta\phi_{\text{bs}}$ defined to be negative/positive when the s-polarisation fringe signal is lagging/leading.

Under the assumption that the piezo acts linearly apart from the turning points of the ramp, the fringe signals can be described by a single frequency and the Pearson correlation coefficient can be used to estimate the phase shift between the two signals. It is defined as the ratio between the covariance of the demeaned signals and the product of their standard deviations \cite{Rodgers1988},
\begin{align}
    r = \frac{\text{cov}(V_{\text{p}},V_{\text{s}})}{\sigma(V_{\text{p}})\sigma(V_{\text{s}})}
    \approx\cos(\Delta\phi_{\text{bs}})
    \,.
\end{align}
This method is insensitive to the sign of the phase shift due to symmetry of the cosine function. To account for multi-frequency components and noise, the result was compared to phase shifts obtained by finding the maximum in the cross-correlation function between both signals, but no significant difference was found. 

\begin{figure}
    \centering
    \includegraphics[width=0.8\linewidth]{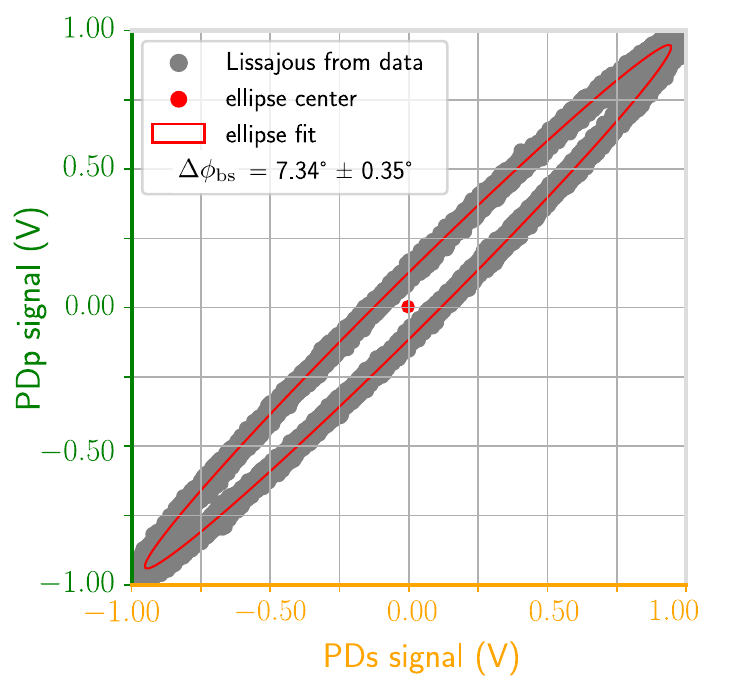}
    \caption{Example of a Lissajous figure from an Optoman sample including fitted ellipse and phase shift from ellipse. The uncertainty was obtained via the standard deviation of the fit parameters from the residuals between fit and data.}
    \label{fig:phi-lissajous}
\end{figure}

To improve the robustness and reliability of the phase shift estimation, especially in the presence of noise, multiple methods were applied. In addition to the Pearson correlation-based cosine inversion and cross-correlation analysis, the geometric ellipse fitting of the Lissajous figure formed by the two fringe signals was used to extract the phase shift. This method additionally provided an intuitive way to infer the sign of the phase shift from the direction of rotation of the Lissajous figure. In the case of the two fringe signals from the scanning Michelson interferometer, the Lissajous figure is a noisy ellipse, see example in fig.~\ref{fig:phi-lissajous}. A numerically stable least square fitting method for ellipses \cite{Halir1998} was used to obtain values for the semi-major $a$ and semi-minor axes $b$ of the normalised ellipse, from which the phase shift can be calculated via
\begin{align}
    \Delta\phi_{\text{bs}}=2\cdot\arctan\left(\frac{b}{a}\right)
    \,.
\end{align}
This method is sensitive to the sign of the phase shift by means of the direction - either clockwise or counterclockwise - in which the points $(V_{\text{p}}(t), V_{\text{s}}(t))$ move along the ellipse as time $t$ progresses. This geometric interpretation of the sign was determined from the data by constructing a velocity vector $(\dot{V}_{\text{p}}(t), \dot{V}_{\text{s}}(t))$ from subsequent data points $\dot{V}_{i}(t)=V_{i}(t+\Delta t)-V_{i}(t)$ ($\Delta t \approx 14\,\text{ns}$) for the datasets of both polarisations individually and by calculating the cross product $X$ between position and velocity vectors,
\begin{align}
    X = V_{\text{s}}\dot{V}_{\text{p}}-V_{\text{p}}\dot{V}_{\text{s}}
    \,.
\end{align}
The sign of the cross product corresponds to the sign of the phase shift with $X<0$ clockwise and s-pol fringes lagging ($\Delta\phi_{\text{bs}}<0)$, and $X>0$ counterclockwise and s-pol fringes leading ($\Delta\phi_{\text{bs}}>0$). 

The uncertainty of the phase shift estimation from the above data analysis methods was quantified using three independent approaches: Monte-Carlo simulation, bootstrap resampling, and analysis of residuals from the ellipse fit. In the Monte-Carlo method, Gaussian noise with a standard deviation of 0.01 was added to the original signals, and the phase was recomputed across 1,000 trials. This process yielded a standard deviation in the phase estimate of approximately $0.01^{\circ}$, providing a measure of statistical uncertainty under the influence of measurement noise. Similarly, bootstrap resampling was used to generate 1,000 synthetic datasets by sampling with replacement from the original time series. The resulting spread in phase estimates yielded a standard deviation close to $0.01^{\circ}$, consistent with the Monte-Carlo result. A third independent estimate of uncertainty was obtained from the ellipse-fitting method by calculating the residuals, defined as the distances from each data point to the closest point on the fitted ellipse. This geometric error metric can be used as a proxy for the precision of the ellipse fit. Phase shift values from variation of ellipse-fitting parameters compatible with the data were found to have an uncertainty of up to $0.35^{\circ}$ for all measurements, which is most likely an overestimation. 

Inconsistencies between repeated measurements for the same sample and between different samples from the same coating run were found to be the largest sources of uncertainty, more than an order of magnitude larger than uncertainties from the data analysis processes detailed above. It is hard to identify the exact cause of these systematic errors, but they can be quantified via the standard deviation of results from different measurements, which is found to be less than $1.5^{\circ}$ for all datasets. Additionally, the phase shift was measured with, respectively, only p- or s-polarised light in the interferometer, turning the polarisation to $45^{\circ}$-linear at the dark port after recombination at the beamsplitter to have equal fringe amplitude on the photodetectors. The average phase shift for these measurements was found to be $-1.75^{\circ}$ without large variations for different beamsplitters and all angles of incidence. This phase shift was understood to be an offset to all measurements. 

\begin{figure}
    \centering
    \includegraphics[width=\linewidth]{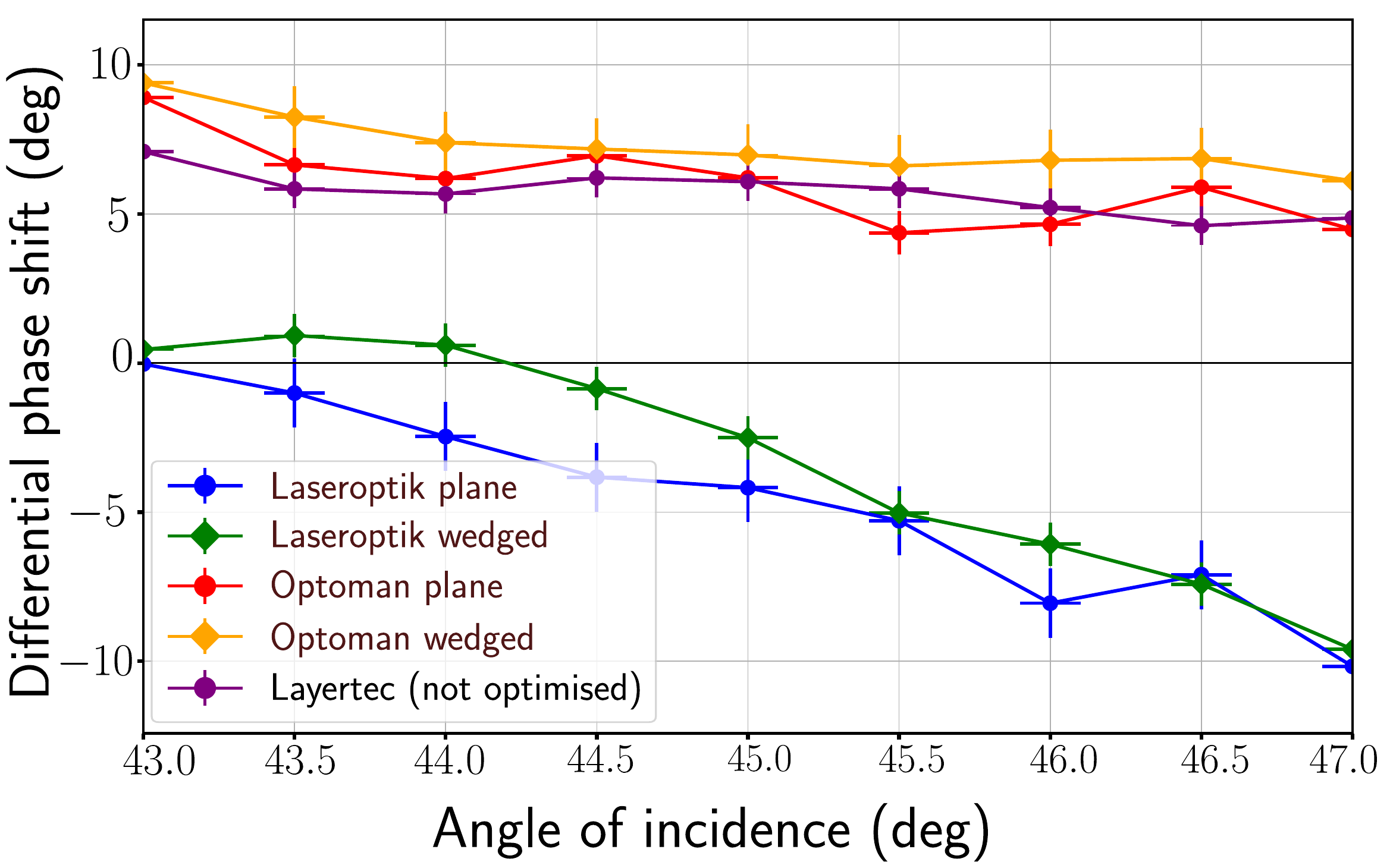}
    \caption{Summary of results from phase shift measurement in the scanning Michelson experiment for an off-the-shelf Layertec coating (purple), and two best-effort coatings from Optoman (orange, red) and Laseroptik (blue, green).}
    \label{fig:phi-result}
\end{figure}

Fig.~\ref{fig:phi-result} shows a summary of the results obtained with the above-mentioned methods. Each curve shows the mean phase shift measured for one type of sample with error bars from the variation between samples on the order of $1.5^{\circ}$ as mentioned above. The datasets for the two custom beamsplitter coatings ordered on best-effort basis from Laseroptik (blue, green) and Optoman (red, orange) with the aim to minimise this phase are compared against measured values for a commercial off-the shelf beamsplitter from Layertec (purple) optimised for s-polarisation. The result for Optoman coatings (orange, red) is more or less independent from the angle of incidence and in the same order of magnitude as for the non-optimised Layertec coating (purple). The Laseroptik coating (green, blue) shows an angle of incidence dependence and reaches a zero-phase shift, but for a smaller angle of incidence compared to the target $45^{\circ}$. This means that a zero-phase shift coating is possible with the Laseroptik design strategy, but needs some optimisation to be reached at the desired orientation of the symmetric beamsplitter. For the Optoman coating, strategies to compensate for the stable phase shift could be investigated. 

\section{Discussion}
In this paper, we analyse two all-polarisation beamsplitter coatings from manufacturers that pursue fundamentally different strategies to achieve similar optical properties for two orthogonal linear polarisation modes. The Laseroptik coating uses a complex design with 22 alternating layers of common materials silica (SiO$_{2}$) and tantala (Ta$_{2}$O$_{5}$) which achieves the desired equal power splitting ratio and zero differential phase, however, at different angles of incidence and not at the target $45^{\circ}$. The complexity of the coating design, which also leads to the significant angle-of-incidence dependence, implies a large number of degrees of freedom that could be used to further optimise this coating design so that both design goals coincide. At the same time, such a design has a high sensitivity to parameter variations during the deposition, so that these have to be tightly monitored and controlled to reach the specifications. 
\\
The Optoman beamsplitter coating utilises a much thinner coating with 5 layers of SiO$_{2}$ and SiO$_{x}$, which feature a much higher refractive index contrast and result in a coating that does not achieve the specifications of equal and close to 50:50 power splitting ratio but $|R_{\text{p}}-R_{\text{s}}|\sim 2\,\%$ and a dark fringe offset of $\Delta\phi_{\text{bs}}\sim 7^{\circ}$. This does not satisfy the specifications at any angle of incidence, but it is robust against angle-of-incidence changes. Therefore, if the design and deposition control can be tuned to reach the specification, the coating would show a large acceptance angle.
\\
Requirements for beamsplitters to be used in speedmeter-like gravitational-wave detectors can be derived from the effect on the sensitivity and control of the interferometer. Ideally, the dual-recycled Michelson interferometer needs to be controlled to be at the dark fringe for both polarisations at the same time. An effective phase shift between the polarisations introduced by the central beamsplitter can therefore be understood as dark fringe offset for one of the polarisations. Depending on the magnitude of this additional degree of freedom, the interferometer could be operated close to dark fringe for both polarisations, i.e. at $\Delta\phi_{\text{bs}}/2$, or the phase shift needs to be controlled, e.g. by means of a phase plate. Alternatively, the temperature-dependence of the differential phase shift could be investigated to explore the possibility of temperature control for this additional degree of freedom. Similar investigations have been conducted for the differential reflection phase between two modes in the context of high-finesse bichromatic optical cavities \cite{Wei2024-1, Wei2024-2}. An optimal working point away from $45^{\circ}$ angle of incidence might also be compensated for with intermediate telescopes, which are foreseen for future gravitational-wave detectors \cite{Rowlinson2021, Srivastava2022} like Cosmic Explorer \cite{CE_Reitze2019, CE_Evans2021} and Einstein Telescope \cite{ET2010, ETscMaggiore2020, ETblue2025}. However, tolerances for this angular deviation depend on the geometrical properties of the detector in question. 
\\
In terms of quantum-noise reduction, a phase shift between both polarisation modes leads to a leakage of position information into the speedmeter readout and therefore an increased back-action noise. Quantification of this effect and derivation of requirements will be subject to future work.
While the dark fringe offset could potentially be compensated by a phase plate or temperature control, the difference in power splitting ratio could be a problem for simultaneous control of the interferometer for both polarisation modes, and needs to be brought closer to 50:50 for both modes independently in future design iterations of the coating.

\section*{Acknowledgements}
The authors would like to thank J\'er\^ome Degallaix (LMA) for useful comments and discussions. Olivier Janssens (Ghent University, Department of Solid State Sciences) is acknowledged for his assistance with operating the JEOL SEM.
Part of this work was supported by the European Research Council (Advanced grant 101019978). J. De Bolle acknowledges the Research Foundation Flanders (FWO) for a scholarship through an FR Grant (No. 11PRJ24N).

\bibliography{bibliography.bib}

\end{document}